\documentclass[aps,amsmath,amssymb,preprint]{revtex4-2}

\usepackage{xr}
\usepackage{lipsum}
\usepackage{graphicx}
\usepackage{dcolumn}
\usepackage{bm}
\usepackage{hyperref}
\usepackage{url}

\usepackage{textcomp}
\usepackage{physics}
\usepackage{multirow}
\usepackage{color, soul} 

\newcommand{\aF} {\alpha^{2}F(\omega)}
\begin{document}

\title{Unraveling the Robust Superconductivity Phenomenon of High-Entropy Alloy}

\author{Adam D. Smith}
\email{smitha20@uab.edu}
\affiliation{Department of Physics, University of Alabama at Birmingham, Birmingham, Alabama 35294, USA}

\author{Wenjun Ding}
\affiliation{Department of Physics, University of Alabama at Birmingham, Birmingham, Alabama 35294, USA}

\author{Yogesh K. Vohra}
\affiliation{Department of Physics, University of Alabama at Birmingham, Birmingham, Alabama 35294, USA}

\author{Cheng-Chien Chen}
\email{chencc@uab.edu}
\affiliation{Department of Physics, University of Alabama at Birmingham, Birmingham, Alabama 35294, USA}

\date{\today}



\maketitle


{\bf
Recent experiments demonstrate a “robust superconductivity phenomenon” in niobium-based alloys, where the superconducting state remains intact and the critical temperature ($\mathbf{T_c}$) is largely unaffected by external pressure well above tens of gigapascal (GPa) into the megabar regime ($\ge \mathbf{100}$ GPa). Motivated by these observations, we perform first-principles electron-phonon calculations for body-centered cubic Nb and NbTi crystals, as well as for special quasi-random structures of Nb$_{\mathbf{0.5}}$Ti$_{\mathbf{0.5}}$ and (NbTa)$_{\mathbf{0.7}}$(HfZrTi)$_{\mathbf{0.3}}$ high-entropy alloy (HEA). The calculations unravel the underlying mechanism of robust superconductivity, stemming from a compensation effect between varying electronic and phonon properties under pressure. The results also reveal how structural and chemical disorders modify the superconducting state. The first-principles $\mathbf{T_c}$ values agree quantitatively with the experiments throughout the entire pressure range under study. Our work thereby paves the way for exploring superconducting HEAs under pressure via advanced first-principles simulations.
}

High-entropy alloys (HEAs) have emerged as a critical research field over the past two decades~\cite{george2019high,zhang2024frontiers,balaji2024development}. These materials consist of five or more elements in a roughly equimolar composition, and they can be stabilized by the contribution of configurational entropy to the Gibbs free energy. HEAs can exhibit the cocktail effect~\cite{hsu2024clarifying}, which results in superior material performance compared to that of the underlying constituent elements. The enhanced structural and electronic properties, including high strength as well as excellent fatigue and corrosion resistance, make HEAs promising for extreme-environment applications~\cite{yusenko2018high,pickering2021high,wang2021high}. 
More recently, it has been reported that body-centered cubic (bcc) niobium-based HEAs of the form (TaNb)$_{1-x}$(HfZrTi)$_x$ become superconducting at approximately $10-15$ K at ambient pressure~\cite{von2016effect,guo2017robust,guo2019record}, and that this superconductivity remains robust or even enhanced when subjected to pressures ranging from tens of gigapascal (GPa) to the megabar regime ($\ge 100$ GPa).
This behavior contrasts with that of pure Nb metal~\cite{struzhkin1997superconducting,ostanin2000calculated,john2004electron}, where the superconducting $T_c$ is suppressed at high pressure.
In fact, this ``robust superconductivity" already occurs in the parent material bcc NbTi alloy~\cite{guo2019record,zhang2020first,jones2025quasiparticle}, as well as in some other HEAs with different crystal symmetries and chemical compositions~\cite{sun2019high, zeng2024recent}.
Understanding the physical mechanism of robust superconductivity may lead to the successful design and discovery of HEA or multi-component materials with higher $T_c$~\cite{wu2024record,ma2025synthesis,ma2025substitution}.

So far, many computational studies have been conducted along this line of effort~\cite{jasiewicz2019pressure,sobota2022superconductivity,sobota2023superconductivity,jasiewicz2023local,huang2020rsavs,ferreira2024ab}, with electron-phonon (el-ph) interactions well established as the primary driving mechanism, according to the Bardeen-Cooper-Schrieffer (BCS) theory~\cite{bardeen1957theory}.
On the other hand, precise calculation of the superconducting $T_c$ in disordered alloy systems is challenging, as the complex interplay between chemical composition, atomic arrangement, and el-ph interactions remains difficult to simulate.
To tackle the challenges, previous studies have utilized the real-space Korringa-Kohn-Rostoker (KKR) method, which can be combined with density functional theory (DFT) for efficient electronic structure calculation. The disorder effect is typically treated within the coherent potential approximation (CPA)~\cite{jasiewicz2019pressure,sobota2022superconductivity,sobota2023superconductivity,jasiewicz2023local} or the virtual crystal approximation (VCA)~\cite{wu2024record}, which essentially averages or interpolates the behaviors of the constituent atoms in the parent materials. These effective-medium studies have provided several insights regarding the pressure evolution of the electronic structures and the overall trend of superconducting $T_c$ with pressure. However, some critical parameters associated with the el-ph interactions may still rely on experimental Debye temperature or related measurements, which thereby limits the predictive capability. Moreover, an effective medium theory may be insufficient to model material-specific conditions. For crystalline systems, it may be preferable to consider a plane-wave method with a supercell approach to simultaneously capture structural and chemical disorder effects. It is also essential to compute the electronic and phonon properties directly from first principles, in order to advance our understanding and achieve quantitative prediction of $T_c$ in HEAs.

In this paper, we perform state-of-the-art plane-wave DFT calculations to investigate the superconducting $T_c$ of NbTi-based HEAs as a function of pressure. 
To elucidate the roles of structural and chemical disorder, we also study elemental Nb metal, as well as NbTi in both the ordered periodic and disordered alloy phases.
The disorder effect is taken into account by the special quasi-random structure (SQS)~\cite{zunger1990special} method, which represents the most random finite-size supercell that approximates a disordered solid solution in the thermodynamic limit.
Experimentally, (TaNb)$_{1-x}$(HfZrTi)$_x$ superconductors have been successfully synthesized for a wide range of $x$ values between $0.2 - 0.84$~\cite{von2016effect,guo2017robust,guo2019record}, and robust superconductivity under pressure was initially reported and characterized in detail for $x=0.33$. Here, we focus on $x =0.3$ to facilitate a manageable-size SQS model of 20 atoms in the simulation. Even so, the corresponding first-principles electron-phonon calculations remain highly nontrivial.
In particular, the phonon spectra and electron-phonon couplings are calculated using advanced Wannier interpolation techniques, as implemented in the EPW~\cite{giustino2007electron} and Quantum Espresso packages~\cite{giannozzi2009quantum,giannozzi2017advanced}, with a very fine Brillouin-zone grid.
Subsequently, the superconducting $T_c$'s are obtained by using the McMillan-Allen-Dynes equations~\cite{mcmillan1968transition, allen1975transition} based on the Eliashberg theory of BCS superconductivity~\cite{eliashberg1960interactions, eliashberg1961temperature, marsiglio2020eliashberg}.
The resulting $T_c$ values from our {\it ab initio} calculations for all the Nb-based materials under study agree quantitatively well with experimental data. Importantly, the calculations reveal that robust superconductivity is caused by a compensation effect, where the electronic density of states, phonon bandwidth, electron-phonon interaction strength, and screened Coulomb potential exhibit distinct behaviors under pressure, collectively maintaining the robustness of superconductivity with a $T_c$ that is insensitive to pressure changes.
Furthermore, our results show that both chemical composition and structural disorder have important tuning effects on the electronic, phonon, and superconducting properties, which in some cases cannot be simply captured by an effective ensemble average method. These results demonstrate the promising predictive capability of using advanced first-principles calculations to model HEA superconductors.

\noindent \textbf{RESULTS AND DISCUSSION}

\begin{figure*}
    \includegraphics[width=1.0\columnwidth]{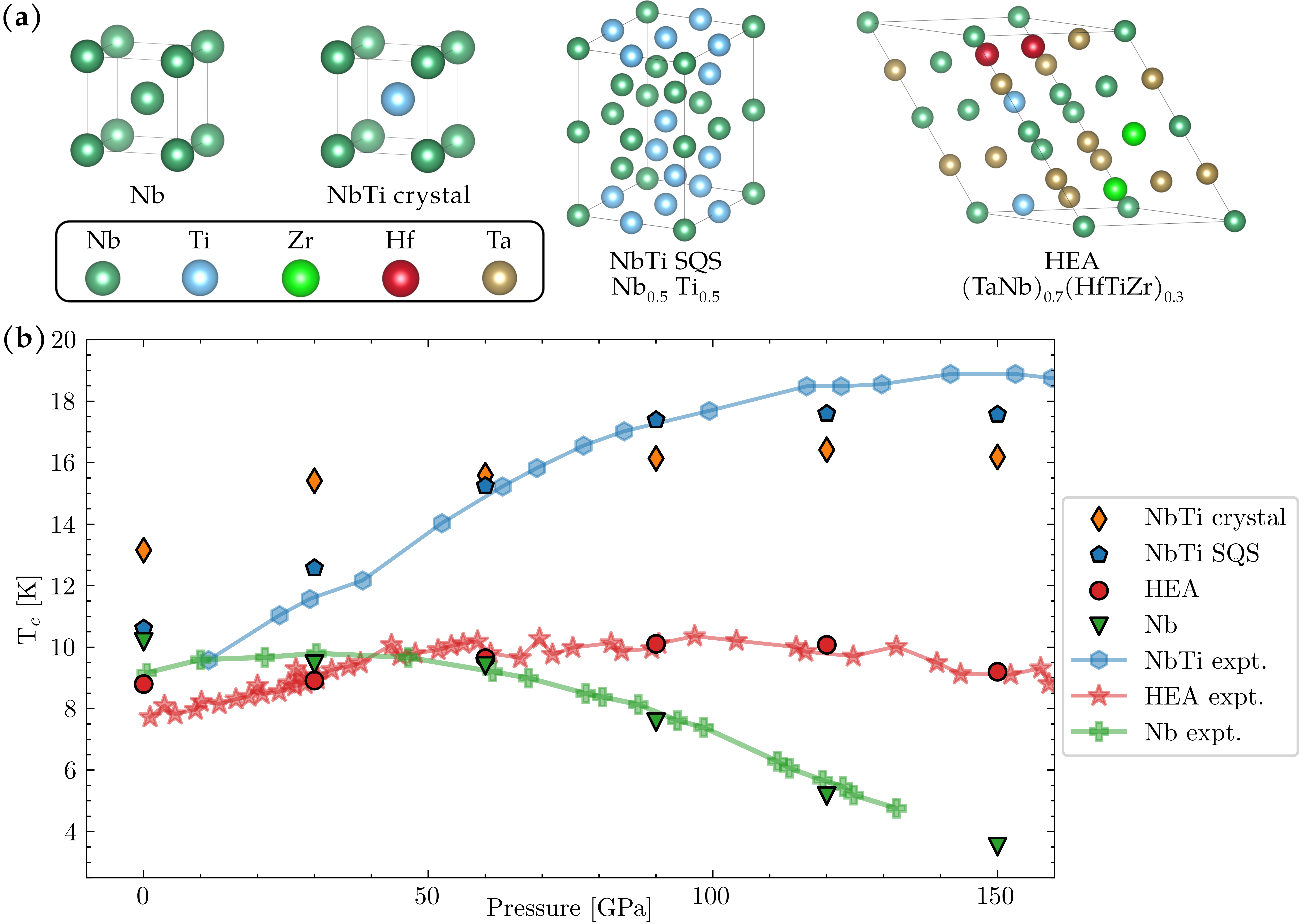}
    \caption{\textbf{Comparison of first-principles calculation and experimental observation on the superconducting $T_c$ in Nb-based materials.} \textbf{(a)} Illustrations of the bcc structures in first-principles calculations, respectively for elemental Nb crystal, NbTi ordered crystal, NbTi alloy (16-atom SQS), and (TaNb)$_{0.7}$(HfZrTi)$_{0.3}$ HEA (20-atom SQS). \textbf{(b)} Superconducting $T_c$ as a function of pressure up to 150 GPa for different Nb-based materials. The computational results are plotted as isolated symbols with black-edged symbols. The experimental $T_c$ values, plotted as solid lines connecting the data points, are based on Refs.~\cite{struzhkin1997superconducting, guo2017robust,guo2019record}. The calculations achieve quantitative theory-experiment agreement. Orange diamonds represent calculations assuming an artificial NbTi ordered crystal, which is dynamically unstable with negative phonon modes at low pressure.}
    \label{Fig1}
\end{figure*}

\noindent \textbf{Superconducting $T_c$ versus Pressure in Nb-based Materials}\\
Figure \ref{Fig1} represents the most significant result of our study: a quantitative comparison of the superconducting $T_c$ values between first-principles calculations and previous experiments in the pressure range of 0 to 150 GPa~\cite{struzhkin1997superconducting, guo2017robust,guo2019record}.
As shown schematically in Fig. \ref{Fig1}(a), four distinct bcc systems are considered in our DFT calculations: elemental Nb, NbTi ordered crystal, NbTi SQS (16 atoms), and (TaNb)$_{0.7}$(HfZrTi)$_{0.3}$ SQS (20 atoms). 
Figure \ref{Fig1}(b) summarizes the experimental $T_c$ values and the corresponding first-principles calculation results. We begin by discussing the general trends of $T_c$ versus pressure in the actual experiments, represented by solid lines connecting the data points in Fig. \ref{Fig1}(b), color-coded as green for elemental Nb metal, blue for NbTi alloy, and red for (TaNb)$_{0.7}$(HfZrTi)$_{0.3}$ HEA.
The experimental $T_c$ of Nb at ambient pressure is approximately 9 K. It has been shown to remain roughly constant at pressures between 0 and 60 GPa, while becoming strongly suppressed beyond 60 GPa.
Afterward, it decreases monotonically with increasing pressure and ultimately reaches $T_c \sim 5$ K at 130 GPa, which represents $\ge 40\%$ reduction in $T_c$ from ambient pressure.
In contrast, the $T_c$ of NbTi alloy increases monotonically with pressure from  $\sim 9$ to 18 K as pressure increases from 0 to 120 GPa; it then remains stable at $\sim 19$ K between 120 and 150 GPa.
In (TaNb)$_{0.7}$(HfZrTi)$_{0.3}$ HEA, $T_c$ increases marginally from around 8 K to 10 K as pressure increases from 0 to 90 GPa. It is stable at 10 K between 90 and 130 GPa, and reduces only slightly to 9.5 K at 150 GPa. The robust $T_c$ behavior against high pressure in NbTi alloy and HEA is markedly different from that of the Nb metal.

We next turn to the first-principles calculation results. The theoretical $T_c$ is computed by the McMillan-Allen-Dynes equations~\cite{mcmillan1968transition, allen1975transition} based on the Eliashberg theory~\cite{eliashberg1960interactions, eliashberg1961temperature, marsiglio2020eliashberg}:
\begin{equation} \label{eqn:ad_tc}
    k_{\rm{B}} T_{\rm{c}} = \frac{\hbar \omega_{log}}{1.2} \mathrm{exp} \left[ - \frac{1.04(1+\lambda)}{\lambda-\mu^{*} (1+0.62\lambda)}\right].
\end{equation}
Here, $\lambda$ is defined as the isotropic electron-phonon coupling strength: 
\begin{equation} \label{eqn:lambda}
    \lambda = 2 \int \frac{\aF}{\omega} d \omega,
\end{equation}
and $\omega_{log}$ is defined as the logarithmic averaged phonon frequency:
\begin{equation} \label{eqn:omega_log}
    \omega_{log} = \mathrm{exp} \left[ \frac{2}{\lambda} \int_{0}^{\infty} d\omega 
    \frac{\aF}{\omega} \log(\omega)\right].
\end{equation}
The key quantity for obtaining $\lambda$ and $\omega_{log}$ is the Eliashberg spectral function $\alpha^2F(\omega)$, which is computed directly from first-principles electron-phonon calculations (see further details in the Methods section).
Finally, $\mu^{*}$ in Eq. (\ref{eqn:ad_tc}) is the the Morel-Anderson Coulomb pseudopotential, which is an effective parameter that describes the effect of screened Coulomb repulsion on superconductivity~\cite{morel1962calculation}.
In the literature, a wide range of $\mu^* \sim 0.1 - 0.3$ has been considered~\cite{giustino2017electron}, and often $\mu^*$ is treated as a fitting parameter by matching the theoretical $T_c$ with the experiment (if available). An analytical expression is also commonly utilized~\cite{bennemann1972theory}:
\begin{equation} \label{eqn:mu*}
    \mu^{*} = K \left( \frac{N_{F}}{N_{F} + 1} \right),
\end{equation}
where $N_{F}$ is the electron density of states (DOS) at the Fermi level ($E_F$). The pre-factor $K$ was originally chosen as an universal value~\cite{bennemann1972theory}.
Here, we consider a material-specific $K$ and constrains it by fitting the calculation results at 6 different pressure points between $0-150$ GPa with the experiments [see Table S1 in the Supplemental Information (SI) for the exact values]. The pressure dependence of $\mu^*$ comes solely from the changes in $N_F$, which is obtained directly from corresponding first-principles calculations. As shown in Fig. \ref{Fig1}(b), the above procedure leads to very good agreement between the theoretical results (isolated symbols with a black edge color) and the experiments (solid lines connecting the data points) for all three Nb-based materials, especially regarding their overall trends of $T_c$ versus pressure and also the exact quantitative $T_c$ values.

\begin{figure} 
    \includegraphics[width=0.6\columnwidth]{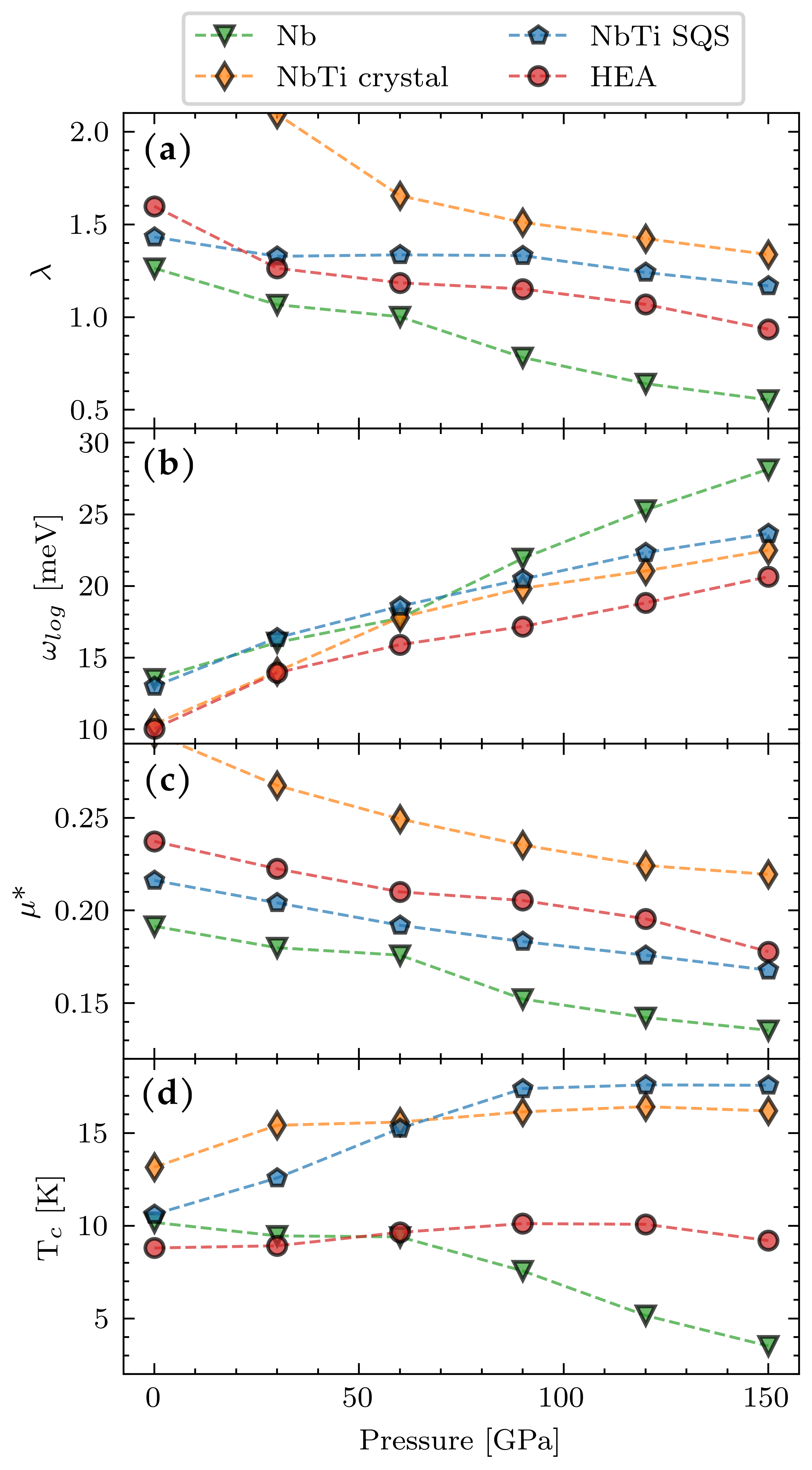}
    \caption{\textbf{Pressure evolution of key parameters that determine the superconducting $T_c$ in first-principles electron-phonon calculations for Nb-based materials.} (a) Electron-phonon coupling strength ($\lambda$), (b) logarithmic averaged phonon frequency ($\omega_{log}$), (c) the Morel-Anderson Coulomb pseudopotential ($\mu^*$), and (d) the resulting superconducting $T_c$. The compensation effect between different parameters collectively leads to a robust superconductivity in NbTi alloy and (TaNb)$_{0.7}$(HfZrTi)$_{0.3}$ HEA at high pressure.}
    \label{Fig2}
\end{figure}

Before we proceed with analyzing how the electronic and phonon properties vary with pressure, a few comments are in order. First, the resulting $\mu^*$ from our calculations lies within the range of 0.15 to 0.25 (as discussed later in Fig. \ref{Fig2}), which is a very reasonable value and completely in line with results in the literature.
Second, even if we simply fit $\mu^*$ using the ambient pressure value (as commonly done in the literature), the overall trends of $T_c$ versus pressure remain the same, with the theoretical $T_c$ in semi-quantitative agreement with the experimental values at high pressure.
Finally, we note that the calculated $T_c$ for the NbTi ordered crystal [orange diamonds in Fig. \ref{Fig1}(b)] deviates substantially from the experimental observation of the NbTi alloy at low pressure, although the calculation does show a robust $T_c$ at high pressure. This deviation is primarily caused by negative phonon modes associated with dynamical lattice instability in the low-pressure NbTi ordered crystal. The negative modes can be removed by the inclusion of anharmonic phonons~\cite{zhang2020first, cucciari2024nbti} or substantial structural disorders, which allow the atoms to relax into a lower-energy configuration in a supercell. This result already highlights the importance of considering alloying effects beyond effective medium theory.


\noindent \textbf{Analysis on the Evolution of $\lambda$, $\omega_{log}$, and $\mu^{*}$ with Pressure}\\
Figure \ref{Fig2} displays first-principles calculations of the key parameters for determining the $T_c$ of BCS superconductors: the electron-phonon coupling strength ($\lambda$), the logarithmic averaged phonon frequency ($\omega_{log}$), and the Morel-Anderson Coulomb pseudopotential ($\mu^*$). We begin by analyzing the general trends across all the Nb-based systems under study.
When pressure increases from 0 to 150 GPa, $\lambda$ decreases by approximately 55\% in Nb, 20\% in NbTi SQS, and 40\% in (TaNb)$_{0.7}$(HfZrTi)$_{0.3}$ HEA, as represented respectively by the green, blue, and red symbols in Fig. \ref{Fig2}(a).
We note that $\lambda$ in the NbTi crystal calculations [orange symbols in Fig. \ref{Fig2}(a)] also decreases with pressure but is unreasonably large in the low-pressure regime, due to low-energy and negative phonon modes associated with structural instability. Overall, the pressure dependence of $\lambda$ is attributed to two factors: a reduction in the electron DOS at the Fermi level ($N_F$), and a hardening of phonon frequencies [see Table S1 in the SI for the exact values]. On general grounds, the electron and phonon bandwidths will increase with pressure due to enhanced charge delocalization and structural hardening caused by compression. Meanwhile, the total electron DOS is conserved yet more spread out with pressure, so $N_F$ is in principle decreased and thereby reduces $\alpha^2F(\omega)$. Moreover, since $\alpha^2F(\omega)$ (and the phonon DOS) are spread out to higher frequencies, $\lambda$ is further decreased by the draining of lower-frequency phonons [due to the $1/\omega$ factor in Eq. (\ref{eqn:lambda})]. Together, these two effects reduce $\lambda$ at high pressure.

On the other hand, Fig. \ref{Fig2}(b) shows that $\omega_{log}$ increases by approximately 1.8 to 2.2 times compared to its ambient-pressure value for all the Nb-based materials at 150 GPa. This increase in $\omega_{log}$ reflects the anticipated phonon hardening due to compression. In contrast, Fig. \ref{Fig2}(c) shows that $\mu^*$ decreases monotonically by approximately 15\% to 25\% for all the systems under study. The reduction in $N_F$ with pressure leads to a reduced $\mu^*$, in accordance with Eq. (\ref{eqn:mu*}). Physically, a reduced effective Coulomb interaction is consistent with the expectation of enhanced charge delocalization and screening effect at high pressure. 

Next, we shift our focus to the pressure dependence of superconducting $T_c$ in different systems, as displayed in Fig. \ref{Fig2}(d). The McMillan-Allen-Dynes formalism [Eq. (\ref{eqn:ad_tc})] dictates that $T_c$ is primarily determined by three key parameters: $\lambda$, $\omega_{log}$, and $\mu^*$. The resulting competition or compensation effect among these factors can lead to either a suppressed $T_c$ or a robust $T_c$ at high pressure. Specifically, in elemental Nb, $T_c$ decreases almost linearly for pressures exceeding 60 GPa. This behavior is closely related to the more substantial decrease in $\lambda$ ($\sim 45\%$ reduction from 60 to 150 GPa), which causes a suppression of $T_c$ from $\sim 9.5$ K at 60 GPa to 3.5 K at 150 GPa. On the other hand, in (TaNb)$_{0.7}$(HfZrTi)$_{0.3}$ HEA, although $\lambda$ also decreases with pressure, it does so at a slower rate. Meanwhile, as pressure increases from 0 to 150 GPa, $\omega_{log}$ of the HEA compound increases from $\sim 10$ meV to 20.65 meV, and its $\mu^*$ decreases from $\sim 0.24$ to 0.18; both effects tend to enhance $T_c$. As a result, the compensation effect among $\lambda$, $\omega_{log}$, and $\mu^{*}$ leads to a robust $T_c$ in (TaNb)$_{0.7}$(HfZrTi)$_{0.3}$ HEA under high pressure. A similar compensation effect also occurs in NbTi alloy, as seen in the NbTi SQS simulation in Fig. \ref{Fig2}(d) (blue pentagons), where its $T_c$ remains robust in the megabar pressure regime ($\ge 100$ GPa). In the NbTi ordered crystal simulation, robust superconductivity is also observed in Fig. \ref{Fig2}(d) (orange diamonds), but the simulated $T_c$ in the low-pressure regime is not applicable due to the existence of negative phonon modes.



\begin{figure*}
    \includegraphics[]
    {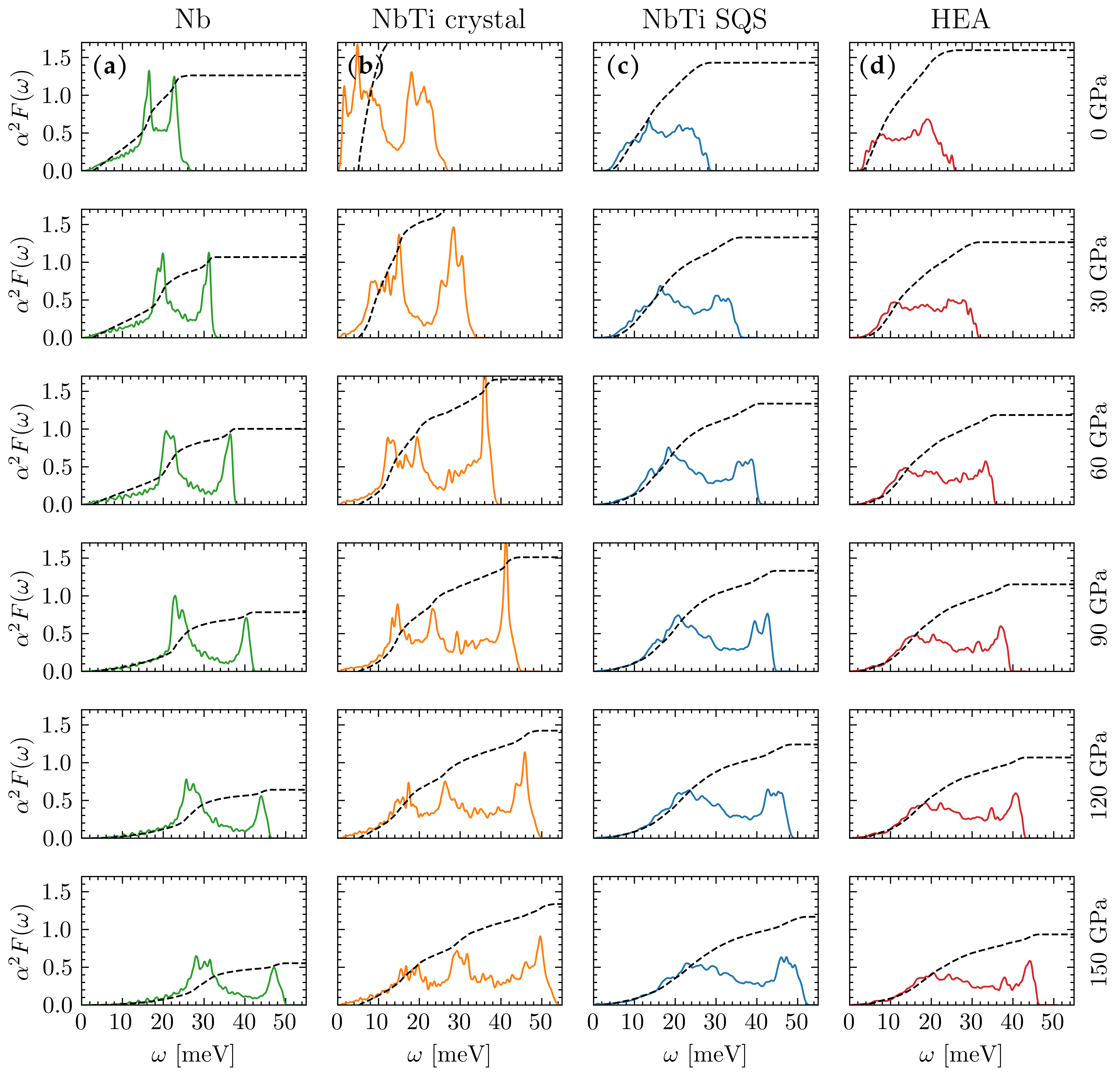}
    \caption{\textbf{Eliashberg spectral functions with pressure in first-principles calculations.} The Eliashberg spectral functions $\alpha^{2}F(\omega)$ computed in the pressure range $0-150$ GPa (with a 30 GPa increment), for \textbf{(a)} Nb metal (green, first column), \textbf{(b)} NbTi crystal (orange, second column), \textbf{(c)} NbTi SQS (blue, third column), and \textbf{(d)} HEA (red, fourth column). The accumulated electron-phonon coupling strengths $\lambda$ based on Eq. (\ref{eqn:lambda}) are plotted in dashed black lines. The progression down each column illustrates how electron-phonon interactions evolve with increasing pressure in each system. The NbTi crystal calculation assumes an artificial ordered structure, which is dynamically unstable with negative phonon modes (not shown) at low pressure.
    }
    \label{Fig3}
\end{figure*}

\noindent \textbf{The Roles of Structural Disorder and Elemental Composition}\\
In the last part of our paper, we further elucidate the impact of structural disorder and elemental composition on the superconducting properties. Figure \ref{Fig3} illustrates the Eliashberg spectral functions $\alpha^2F(\omega)$ computed in the pressure range of $0-150$ GPa (with a pressure increment of 30 GPa) for all the Nb-based systems under study.
We note that the computational cost is significant for the 16- and 20-atom SQS models, as the calculations are performed on very fine momentum-space sampling grids using Wannier interpolation.
Overall, applying pressure increases the bandwidths of $\alpha^2F(\omega)$ (and the phonon DOS, as shown in the SI), with the spectral peaks shifted to higher energies due to pressure-enhanced structural hardening.
To reveal the effect of structural disorder, we compare the calculations using the NbTi ordered crystal and the NbTi SQS model, as shown respectively in Figs. \ref{Fig3}(b) and \ref{Fig3}(c).
Notably, in the low-pressure regime ($0–30$ GPa), the NbTi crystal exhibits substantial low-energy spectral weight below 10 meV, while the NbTi SQS model lacks these low-energy modes. The presence of low-energy peaks is caused by soft phonons (or even negative phonon modes in the 0 GPa case) in the NbTi crystal, which indicates dynamical instability and proximity to a structural transition. In comparison, the NbTi SQS model remains dynamically stable across all pressures studied. In the higher-pressure range ($60-150$ GPa), the NbTi crystal also becomes stable, and the impact of structural disorder as simulated by the NbTi SQS model essentially reduces the high-energy sharp peaks and effectively smooths the spectra. These results are consistent with a spectral averaging effect in disordered alloys.
As discussed previously in Fig. \ref{Fig1}(b), a quantitative agreement between first-principles theory and the experimental $T_c$ of NbTi alloy is achieved for all pressures only in the SQS model. Considering dynamical stability as well, our findings confirm the critical importance of explicitly incorporating structural disorder effects to accurately capture the electron-phonon coupling and superconducting behaviors of alloy superconductors.

Finally, we discuss the influence of elemental composition. In several aspects, $\alpha^2F(\omega)$ of (TaNb)$_{0.7}$(HfZrTi)$_{0.3}$ HEA [Fig. \ref{Fig3}(d)] closely resembles that of NbTi SQS [Fig. \ref{Fig3}(c)]. Both alloy systems exhibit a similar spectral profile across the entire pressure range, with spectral peaks effectively smoothed and broadened by structural disorder effects. Some noticeable differences include a slightly reduced bandwidth and suppressed peak intensities around 20 meV in the HEA, compared to the NbTi SQS model. These differences arise from varying chemical compositions of the two systems. 
As shown above in Fig. \ref{Fig1}(b), while the $T_c$ in both systems is comparable at ambient pressure, the NbTi $T_c$ is nearly twice that of (TaNb)$_{0.7}$(HfZrTi)$_{0.3}$ at 150 GPa. This enhancement in $T_c$ primarily arises from a more favorable pressure dependence of electron-phonon coupling $\lambda$ in the NbTi SQS, as  $\omega_{log}$ and $\mu^*$ are comparable across the systems (see Fig. \ref{Fig2}). Under the isotropic approximation, Eq. (\ref{eqn:lambda}) can be written as $\lambda = N_F \langle g^2 \rangle/M\langle \omega^2 \rangle$~\cite{mcmillan1968transition, hopfield1969angular}. Here, $\langle g^2 \rangle$ is the square of the electron-phonon coupling matrix elements averaged over the Fermi surface, $M$ is the atomic mass, and $\langle \omega^2 \rangle$ is the averaged phonon frequency square. 
A higher content of the lighter element Ti can contribute to a smaller average atomic mass, while also resulting in a larger phonon bandwidth.
External pressure would reduce $N_F$ and enhance $\langle \omega^2 \rangle$ simultaneously, both leading to a decreased $\lambda$.
Overall, $\lambda$ and $T_c$ can depend on a delicate balance of factors.

Further insight is gained by comparing the projected phonon densities of states (DOS). In NbTi SQS (Fig. S3 in SI), the low-energy peaks in $\alpha^{2}F(\omega)$ mainly originate from Nb (and to a lesser extent, Ti), which proportionately enhance $\lambda$ [due to the $1/\omega$ factor in Eq. (\ref{eqn:lambda})]. 
In contrast, in the HEA system (Fig. S4 in SI), the majority of low-energy phonon peaks stem from Ta, followed by Nb at slightly higher energy. Pressure tends to increase the phonon bandwidth and spread out the spectral weights, but its effects can be non-linear and vary between different systems.
For example, in the evolution of $\omega_{log}$ beyond 60 GPa [Fig. \ref{Fig2}(a)], elemental Nb's phonons harden significantly faster than those in the NbTi SQS and HEA, whose $\omega_{log}$ evolve more consistently with pressure. Consequently, both $\lambda$ and $T_c$ decrease significantly after 60 GPa in elemental Nb. These trends underscore the beneficial role of chemical and/or structural disorder in mitigating the rapid deterioration of electron-phonon coupling and $T_c$ values under extreme pressure conditions. Therefore, comprehensive optimization of both compositional and structural parameters, as well as their pressure dependence, is imperative for achieving a higher $T_c$ in Nb-based alloy superconductors.

To conclude, we have conducted a comprehensive first-principles investigation of phonon-mediated superconductivity in elemental Nb, ordered NbTi, disordered NbTi alloy, and (TaNb)$_{0.7}$(HfZrTi)$_{0.3}$ high-entropy alloy under pressure. Our calculations reveal that robust superconductivity in alloy systems originates from delicate compensation
effects among the electronic density of states, phonon hardening, electron-phonon
coupling strength, and screened Coulomb repulsion. Structural and chemical disorders are shown to significantly alter the phonon landscape and the resulting Eliashberg spectral functions, enabling enhanced or sustained $T_c$ into the megabar regime ($\ge 100$ GPa). Our results further highlight the importance of incorporating realistic disordered structure models to accurately capture the behaviors of alloy superconductors under pressure. 
While high-entropy materials present several challenges in studying their properties, they also offer promising opportunities through engineering compositions and structural effects to achieve robust superconductivity and potentially higher $T_c$.
Finally, we note that despite the success in achieving quantitative theory-experiment agreements, our methodologies may be limited in terms of the supercell size, thereby restricting the possible stoichiometry and composition that can be directly simulated. Therefore, benchmarking our first-principles results with other advanced methods and applying our methodologies to other high-entropy superconductors with varying crystal symmetries and chemical compositions will be both important areas of future research.

\noindent \textbf{METHODS}

\noindent \textbf{Density Functional Theory Calculations}\\
Our first-principles calculations utilize \textsc{Quantum ESPRESSO} (version 7.3.1)~\cite{giannozzi2009quantum, giannozzi2017advanced}, an open-source, plane-wave, pseudopotental-based density functional theory (DFT) software.
We employ norm-conserving, scalar-relativistic pseudopotentials from Pseudo-Dojo (version 0.5)~\cite{van2018pseudodojo}, generated using the Perdew-Burke-Ernzerhof (PBE) exchange-correlation functional~\cite{perdew1996generalized}.
A plane-wave cutoff energy of 85 Ry is used, which is sufficient to converge the total energies to within $10^{-6}$ Ry/atom in all the primitive unit cells under study. A Gaussian smearing of 0.015 Ry is also applied to account for partial occupancies near the Fermi level.
All systems studied are relaxed under external pressures ranging from 0 to 150 GPa in 30 GPa increments. The relaxation criteria are set to ensure convergence of the total energies to within $10^{-9}$ Ry/atom and forces to less than $10^{-6}$ Ry/bohr per atom. We adopt a 0.15 \AA$^{-1}$ $\Gamma$-centered Monkhorst-Pack grid for sampling the Brillioun zone. After structural relaxation, self-consistent field calculations are conducted to produce the ground-state charge density and Bloch wavefunctions for successive density functional perturbation theory (DFPT) and electron-phonon coupling calculations.

\noindent \textbf{Special Quasi-Random Structures}\\
To account for disorder and alloying effects, we employ the special quasi-random structure (SQS)~\cite{zunger1990special}, which provides the best periodic supercell approximation to a truly disordered state. In particular, the highly efficient Monte Carlo SQS (\textsc{mcsqs}) \cite{van2013efficient} code implemented in the \textsc{Alloy Theoretic Automated Toolkit (ATAT)}~\cite{van2002alloy} is used. \textsc{mcsqs}, based on simulated annealing with an objective function that maximizes the number of matched correlation functions, optimizes both the shape of the supercell and the occupation of atomic sites.
SQS clusters are generated using a two-atom body-centered cubic (bcc) unit cell, partially occupied by Nb and Ti in equal proportions for the Nb$_{0.5}$Ti$_{0.5}$ alloy, and by $\frac{7}{20}$Ta, $\frac{7}{20}$Nb, $\frac{2}{20}$Hf, $\frac{2}{20}$Zr, and $\frac{2}{20}$Ti for the (TaNb)$_{0.7}$(HfZrTi)$_{0.3}$ high-entropy alloy (HEA). The algorithm was run until the resulting SQS structure remained unchanged for 24 hours.

\noindent \textbf{Density-Functional Perturbation Theory and Electron-Phonon Coupling}\\
DFPT calculations are conducted to produce the dynamical matrices, force constants, and phonon spectra for all the systems under study. In each DFPT calculation, the threshold for self-consistency is set to 10$^{-17}$ Ry$^2$ (which represents the squared norm of the change in the first-order potential). Dynamical matrices are calculated on the irreducible Brillouin zone with $\mathbf{q}$ grids of $8\times 8 \times 8$, $7 \times 7 \times 7$, $3 \times 3 \times 2$, and $3 \times 3 \times2$, respectively for Nb, NbTi crystal, NbTi SQS, and the HEA system. The electron-phonon coupling matrix elements are computed with the \textsc{EPW} code~\cite{giustino2007electron}, which uses Maximally Localized Wannier Functions (MLWFs)~\cite{marzari2012maximally} generated by the \textsc{Wannier90} package~\cite{mostofi2014updated} and the dynamical matrices produced by DFPT to interpolate the electron-phonon matrix elements to high-resolution $\mathbf{k}$ and $\mathbf{q}$ grids.
We use atomically centered $d$ orbitals as initial projections for \textsc{Wannier90}.
After careful convergence tests, $90 \times 90 \times 90$ $\mathbf{k}$ and $30 \times 30 \times 30$ $\mathbf{q}$ grids are chosen for the Nb and NbTi crystals; $24 \times 24 \times 16$ $\mathbf{k}$ and $24 \times 24 \times 16$ $\mathbf{q}$ grids are adopted for the Nb$_{0.5}$Ti$_{0.5}$ SQS and (TaNb)$_{0.7}$(HfZrTi)$_{0.3}$ HEA systems. 

\noindent \textbf{DATA AVAILABILITY}\\
The main input and output files of our density functional theory and electron-phonon calculations can be found at {\url {https://github.com/condmatr/Robust_Superconductivity}}.

\noindent \textbf{ACKNOWLEDGMENTS}\\
The research is supported by the U.S. National Science Foundation (NSF) Award No. DMR-2310526. W.D. and C.-C.C. also acknowledge support from NSF Award No. DMR2142801. The calculations utilized the Frontera computing system at the Texas Advanced Computing Center, which is made possible by NSF Award No. OAC-1818253.

\noindent \textbf{AUTHOR CONTRIBUTIONS}\\
A.D.S. performed the calculations and analyzed the data; A.D.S., W.D., and C.-C.C. prepared the first draft of the manuscript; all
authors contributed to discussion of the results and revision of the manuscript; Y.K.V. and C.-C.C. initiated and coordinated the project.

\noindent \textbf{COMPETING INTERESTS}\\
The authors declare no competing interests.


\bibliography{main}

\end{document}


\title{Supplemental Information:\\
Unraveling the Robust Superconductivity Phenomenon of
High-Entropy Alloy} 	

\author{Adam D. Smith}
\email{smitha20@uab.edu}
\affiliation{Department of Physics, University of Alabama at Birmingham, Birmingham, Alabama 35294, USA}

\author{Wenjun Ding}
\affiliation{Department of Physics, University of Alabama at Birmingham, Birmingham, Alabama 35294, USA}

\author{Yogesh K. Vohra}
\affiliation{Department of Physics, University of Alabama at Birmingham, Birmingham, Alabama 35294, USA}

\author{Cheng-Chien Chen}
\email{chencc@uab.edu}
\affiliation{Department of Physics, University of Alabama at Birmingham, Birmingham, Alabama 35294, USA}
\date{\today}

\maketitle

\vfill\null

\pagebreak

\begin{table}[h]
\setlength{\tabcolsep}{5pt} 
\begin{center}
\caption{
Summary of first-principles numerical data used to determine the Allen-Dynes superconducting transition temperature $T_{c,AD}$ (in units of Kelvin) under different pressures. The fitted prefactor $K$, used to evaluate the Morel-Anderson pseudopotential $\mu^*$ as discussed in the main text, is provided in parentheses under each material.
Pressure is given in units of GPa. $N_{F}$ is the normalized electron density of states at the Fermi level, with units of states/$\textrm{eV}\cdot\textrm{atom}$. The logarithmic average of phonon frequency $\omega_{log}$ is given in units of meV. $\lambda$ and $\mu^{*}$ are unitless.}

\begin{tabular}{| c | c | c | c | c | c | c |}
\hline

System & Pressure & $N_{F}$ & $\omega_{log}$  & $\lambda$ & $\mu^{*}$ & $T_{c,AD}$ \\ 
\hline
\bottomrule[1.25pt]
\multirow{6}{*}{\makecell{Nb \\ ($K=0.325$)}}
& 0 & 1.44 & 13.52 & 1.26 & 0.19 & 10.18 \\ \cline{2-7}
& 30 & 1.24 & 16.06 & 1.07 & 0.18 & 9.45 \\ \cline{2-7}
& 60 & 1.18 & 17.75 & 1.00 & 0.18 & 9.40 \\ \cline{2-7}
& 90 & 0.88 & 21.92 & 0.78 & 0.15 & 7.56 \\ \cline{2-7}
& 120 & 0.78 & 25.30 & 0.64 & 0.14 & 5.16 \\ \cline{2-7}
& 150 & 0.71 & 28.15 & 0.55 & 0.14 & 3.51 \\ \cline{2-7}
\hline
\bottomrule[1.25pt]

\multirow{6}{*}{\makecell{NbTi\\Crystal \\ ($K=0.431$)}}
& 0 & 2.17 & 10.36 & 2.65 & 0.30 & 13.16 \\ \cline{2-7}
& 30 & 1.64 & 14.03 & 2.09 & 0.27 & 15.42 \\ \cline{2-7}
& 60 & 1.37 & 17.81 & 1.65 & 0.25 & 15.59 \\ \cline{2-7}
& 90 & 1.20 & 19.84 & 1.51 & 0.24 & 16.14 \\ \cline{2-7}
& 120 & 1.08 & 21.05 & 1.42 & 0.22 & 16.42 \\ \cline{2-7}
& 150 & 1.04 & 22.49 & 1.34 & 0.22 & 16.19 \\ \cline{2-7}
\hline
\bottomrule[1.25pt]

\multirow{6}{*}{\makecell{Nb$_{0.5}$Ti$_{0.5}$ \\ SQS \\ ($K=0.348$)}}
& 0 & 1.64 & 12.99 & 1.43 & 0.22 & 10.61 \\ \cline{2-7}
& 30 & 1.42 & 16.38 & 1.33 & 0.20 & 12.57 \\ \cline{2-7}
& 60 & 1.23 & 18.59 & 1.34 & 0.19 & 15.25 \\ \cline{2-7}
& 90 & 1.11 & 20.48 & 1.33 & 0.18 & 17.39 \\ \cline{2-7}
& 120 & 1.02 & 22.33 & 1.24 & 0.18 & 17.59 \\ \cline{2-7}
& 150 & 0.93 & 23.65 & 1.17 & 0.17 & 17.57 \\ \cline{2-7}
\hline
\bottomrule[1.25pt]

\multirow{6}{*}{\makecell{(TaNb)$_{0.7}$\\(HfZrTi)$_{0.3}$ \\ SQS \\ ($K=0.384$)}}
& 0 & 1.62 & 10.04 & 1.60 & 0.24 & 8.80 \\ \cline{2-7}
& 30 & 1.38 & 13.94 & 1.26 & 0.22 & 8.92 \\ \cline{2-7}
& 60 & 1.21 & 15.90 & 1.19 & 0.21 & 9.64 \\ \cline{2-7}
& 90 & 1.15 & 17.17 & 1.15 & 0.21 & 10.11 \\ \cline{2-7}
& 120 & 1.04 & 18.82 & 1.07 & 0.20 & 10.08 \\ \cline{2-7}
& 150 & 0.86 & 20.65 & 0.93 & 0.18 & 9.20 \\ \cline{2-7}
\hline
\bottomrule[1.25pt]

\end{tabular}
\end{center}
\end{table}

\begin{figure*}
    \includegraphics[]{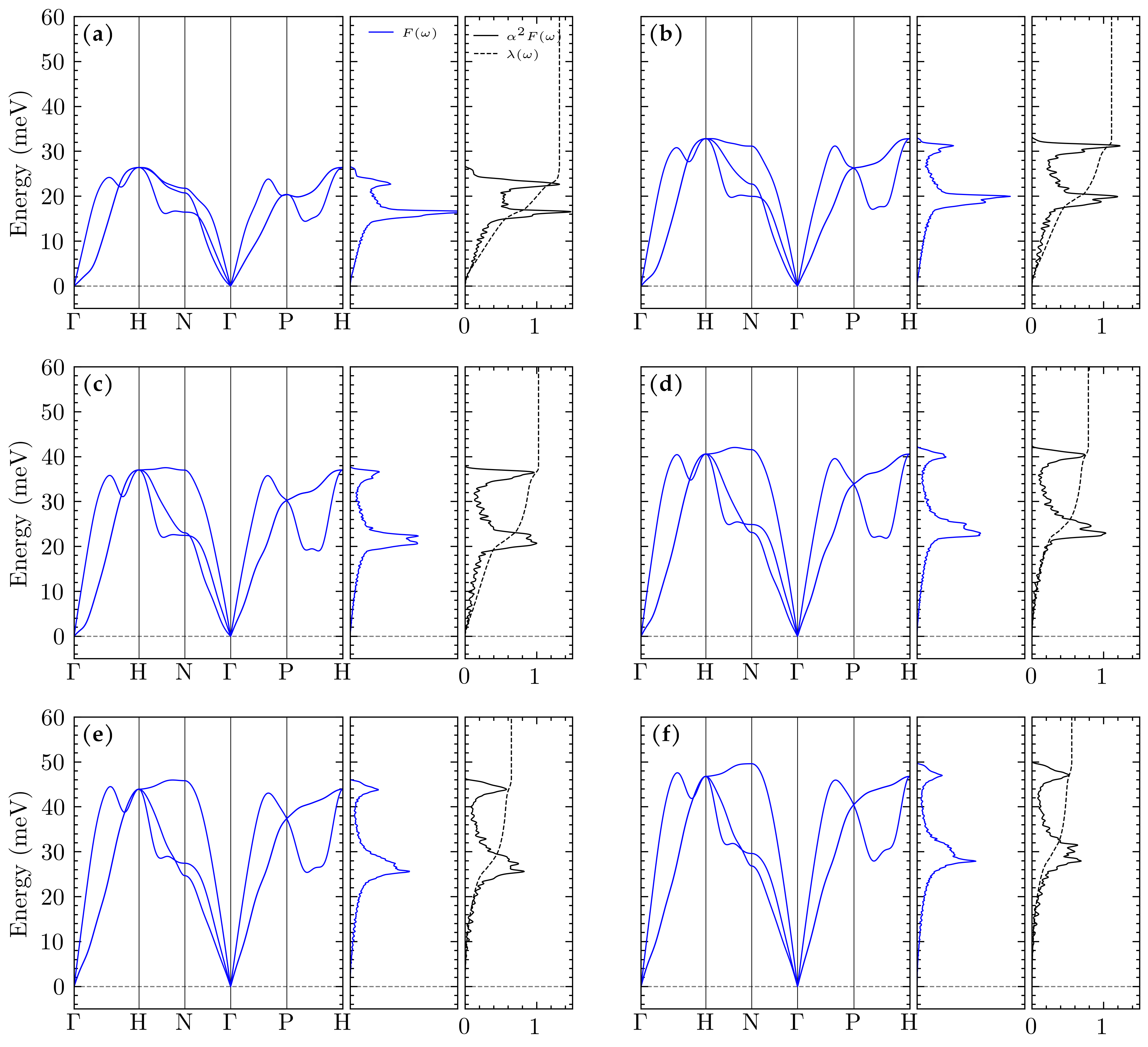}
    \caption{\textbf{Phonon dispersions, phonon density of states (DOS), and the Eliashberg spectral functions for body-centered cubic Niobium (Nb) at pressures 0 to 150 GPa in 30 GPa increments (panels \textbf{(a)}-\textbf{(f)}).}
    In each panel, the left subfigure shows the phonon dispersion; the middle shows the total phonon DOS $F(\omega)$; the right presents the Eliashberg spectral function $\alpha^{2}F(\omega)$ (solid line) and the cumulative electron-phonon coupling $\lambda(\omega)$ (dashed line). The high-symmetry Brillouin zone paths are based on a crystal unit cell of space group $\mathbf{Im\Bar{3}m}$.}
    \label{SIFig1}
\end{figure*}

\begin{figure*}
    \includegraphics[]{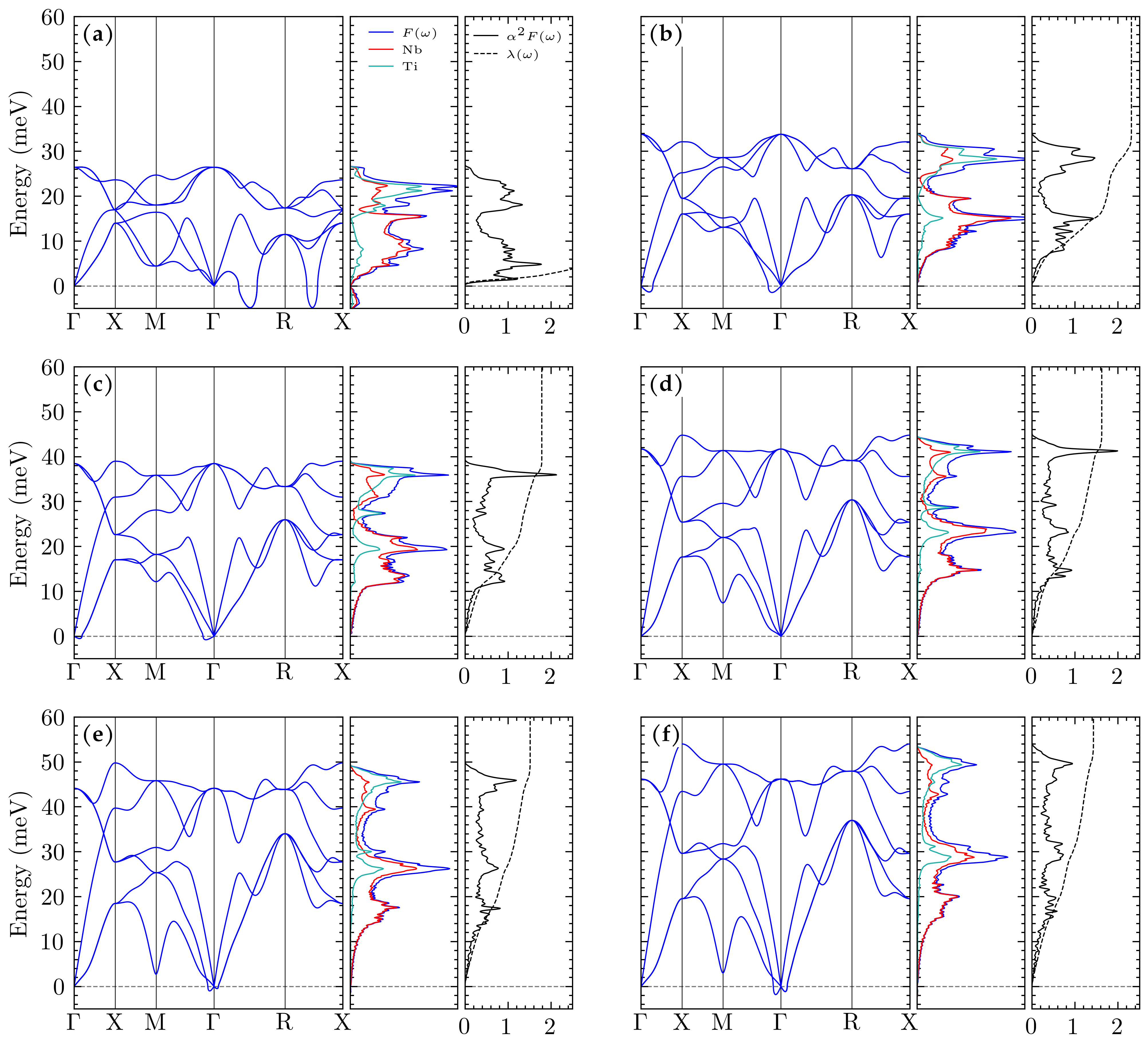}
    \caption{\textbf{Phonon dispersions, phonon density of states (DOS), and the Eliashberg spectral functions for body-centered cubic Niobium-Titanium (NbTi) at pressures 0 to 150 GPa in 30 GPa increments (panels \textbf{(a)}-\textbf{(f)}).} In each panel, the left subfigure shows the phonon dispersion; the middle shows the total phonon DOS $F(\omega)$ and the atomically projected phonon DOS; the right presents the Eliashberg spectral function $\alpha^{2}F(\omega)$ (solid line) and the cumulative electron-phonon coupling $\lambda(\omega)$ (dashed line). The high-symmetry Brillouin zone paths are based on a crystal unit cell of space group $\mathbf{Pm\Bar{3}m}$.}
    \label{SIFig2}
\end{figure*}

\begin{figure*}
    \includegraphics[]{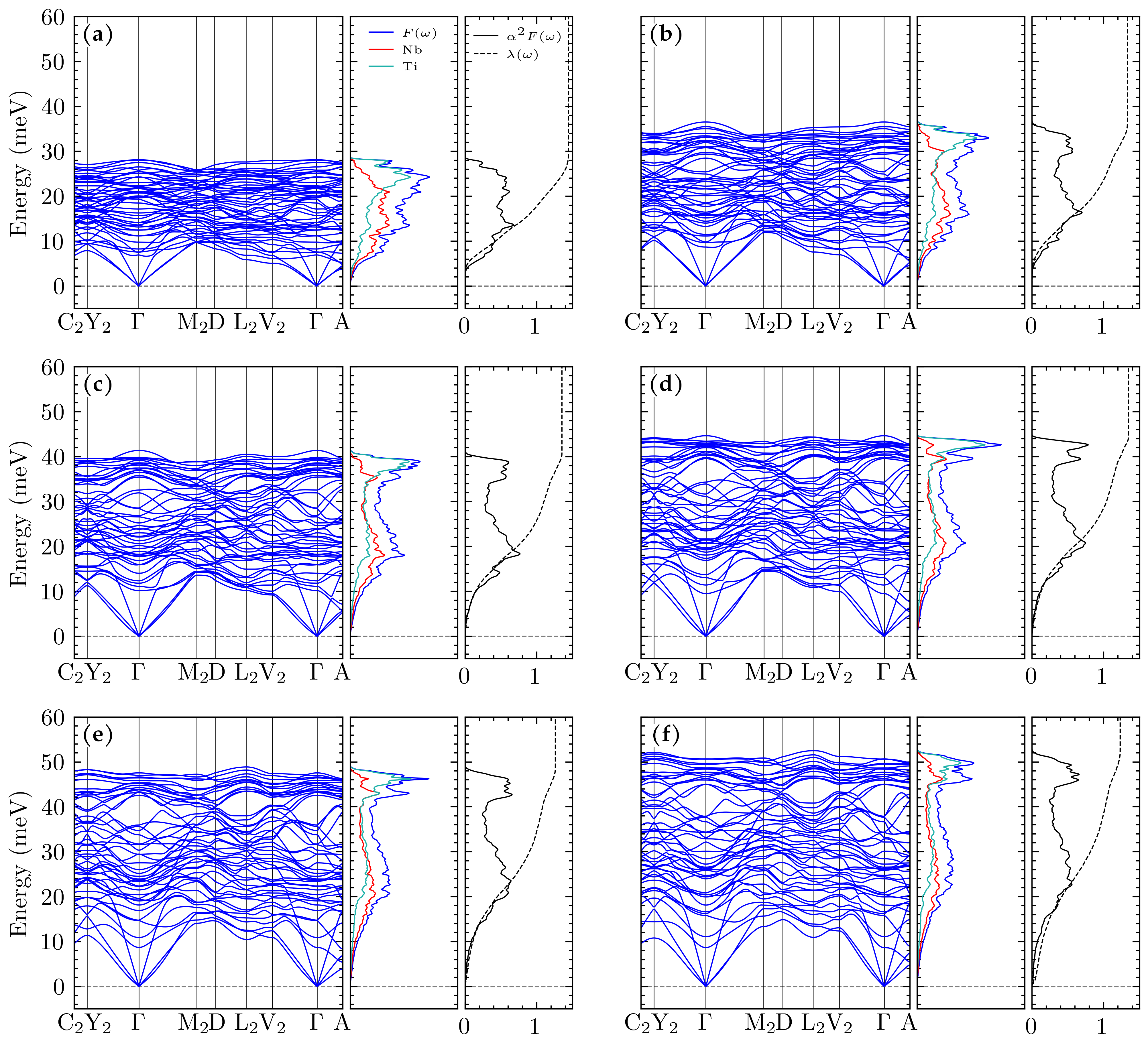}
    \caption{\textbf{Phonon dispersions, phonon density of states (DOS), and the Eliashberg spectral functions for body-centered cubic Nb$_{0.5}$Ti$_{0.5}$ alloy at pressures 0 to 150 GPa in 30 GPa increments (panels \textbf{(a)}-\textbf{(f)}).} In each panel, the left subfigure shows the phonon dispersion; the middle shows the total phonon DOS $F(\omega)$ and the atomically projected phonon DOS; the right presents the Eliashberg spectral function $\alpha^{2}F(\omega)$ (solid line) and the cumulative electron-phonon coupling $\lambda(\omega)$ (dashed line). The high-symmetry Brillouin zone paths are based on a special quasi-random structure unit cell of space group $\mathbf{Cm}$.}
    \label{SIFig3}
\end{figure*}

\begin{figure*}
    \includegraphics[]{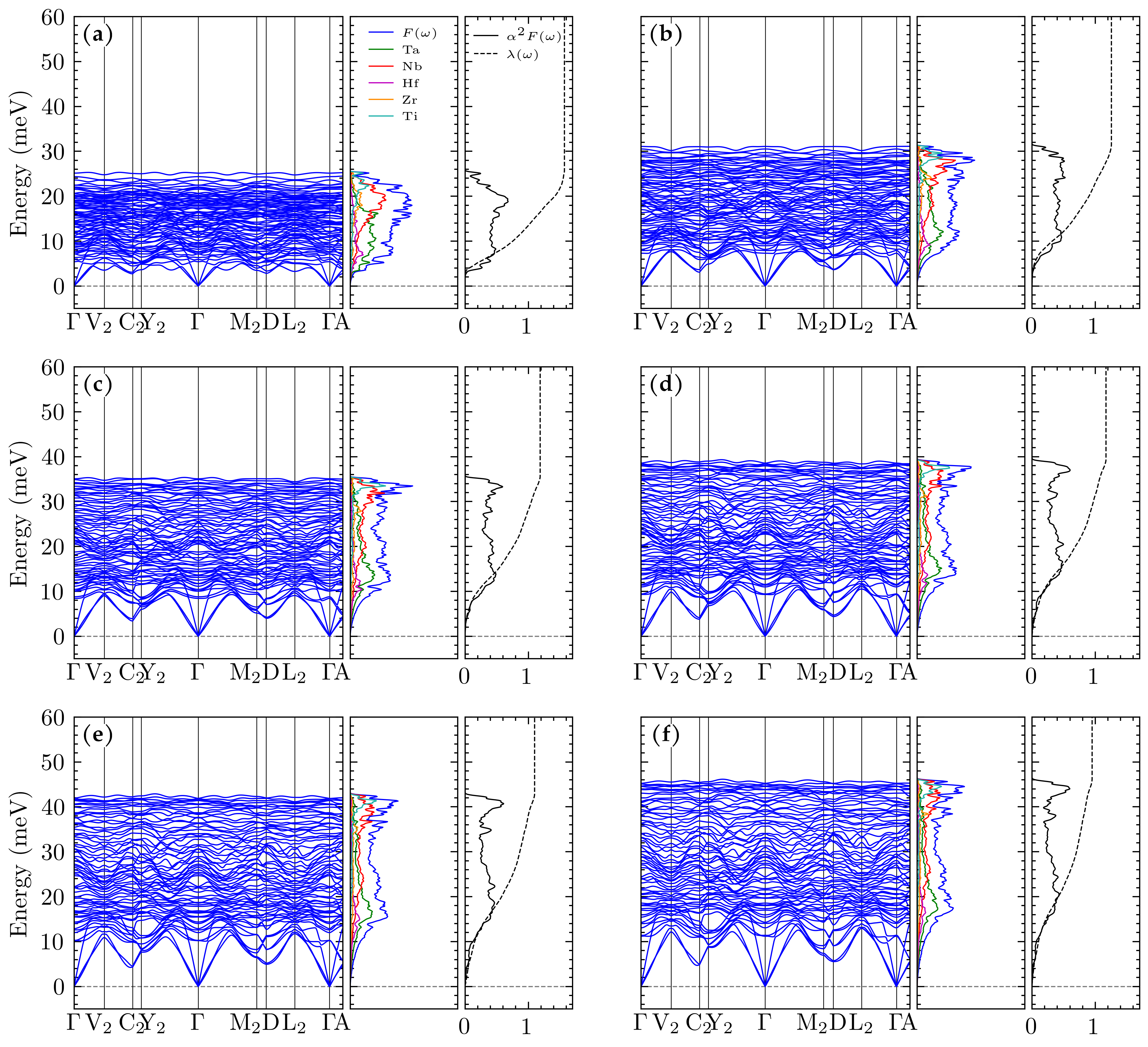}
    \caption{\textbf{Phonon dispersions, phonon density of states (DOS), and the Eliashberg spectral functions for body-centered cubic (TaNb)$_{0.7}$(HfZrTi)$_{0.3}$ high-entropy alloy at pressures 0 to 150 GPa in 30 GPa increments (panels \textbf{(a)}-\textbf{(f)}).} In each panel, the left subfigure shows the phonon dispersion; the middle shows the total phonon DOS $F(\omega)$ and the atomically projected phonon DOS; the right presents the Eliashberg spectral function $\alpha^{2}F(\omega)$ (solid line) and the cumulative electron-phonon coupling $\lambda(\omega)$ (dashed line). The high-symmetry Brillouin zone paths are based on a special quasi-random structure unit cell of space group $\mathbf{Cm}$.}
    \label{SIFig4}
\end{figure*}